\documentclass[aps,twocolumn,amsfonts,amssymb,amsmath,showpacs]{revtex4}
\usepackage{epsfig, graphicx, color, amsmath, amsthm, amssymb}
\input{qcircuit}
\newcommand{\targup}{*{\xy{<0em,0em>*{} \ar @{ - } +<.4em,0em> \ar @{ - } -<.4em,0em> \ar @{ - } +<0em,.4em> \ar @{ - } -<0em,.3em>},*+<.8em>\frm{o}\endxy} \qw}
\renewcommand{\targ}{\targup}
\usepackage{diagrams}
\def\ket #1{\vert #1\rangle}
\def\bra #1{\langle #1\vert}
\def\ketbra #1#2{\ket{#1}\!\bra{#2}}

\DeclareMathOperator{\tr}{Tr}
\newcommand{\beq}{\begin{equation}}
\newcommand{\eeq}{\end{equation}}
\newcommand{\binomial}[2]{\ensuremath{\left(\begin{smallmatrix}#1 \\ #2 \end{smallmatrix}\right)}}

\newcommand{\comment}[1]{\emph{\color{red}Comment:\color{black} #1}} 
\newlength{\commentslength}
\newcommand{\comments}[1]{
\hspace{-2\parindent}
\addtolength{\commentslength}{-\commentslength}
\addtolength{\commentslength}{\linewidth}
\addtolength{\commentslength}{-\parindent}
\fcolorbox{red}{white}{\smallskip\begin{minipage}[c]{\commentslength}
\emph{Comments:}\begin{itemize}#1\end{itemize}\end{minipage}}\bigskip
}

\newtheorem{theorem}{Theorem}  

\newtheoremstyle{note}{}{}{\slshape}{}{\bfseries}{.}{ }{}
\theoremstyle{note}

\newtheorem{definition}{Definition}

\renewcommand{\comment}[1]{}\renewcommand{\comments}[1]{}

\DeclareMathOperator{\poly}{\operatorname{poly}}

\begin{document}

\author{Ben W. Reichardt}
\email{breic@cs.berkeley.edu}
\affiliation{{EECS} Department, Computer Science Division, University of California, Berkeley, California 94720}


\title{Fault-tolerance threshold for a distance-three quantum code}

\begin{abstract}
The quantum error threshold is the highest (model-dependent) noise rate which we can tolerate and still quantum-compute to arbitrary accuracy.  
Although noise thresholds are frequently estimated for the Steane seven-qubit, distance-three quantum code, 
there has been no proof that a constant threshold even exists for distance-three codes.  
We prove the existence of a constant threshold.
The proven threshold is well below estimates, based on simulations and analytic models, of the true threshold, but at least it is now known to be positive.  
\end{abstract}

\maketitle

\section{Introduction} \label{s:introduction}

Quantum operations are inherently noisy, so the development of fault-tolerance techniques is an essential part of progress toward a quantum computer.  
A quantum circuit with $N$ gates can only a priori tolerate $O(1/N)$ error per gate.  
In 1996, Shor showed how to tolerate $O(1/\poly(\log N))$ error by encoding each qubit into a $\poly(\log N)$-sized quantum error-correcting code, then implementing each gate of the desired quantum circuit directly on the encoded qubits, alternating computation and error-correction steps \cite{Shor96}.  Even though the corrections themselves are imperfect, noise overall remains under control -- the scheme is ``fault-tolerant."

Several groups \cite{AharonovBenOr99,KnillLaflammeZurekProcRSocLondA98,Kitaev96b,GottesmanEvslinKakadePreskill98} independently realized that by instead using a constant-sized quantum error-correcting code repeatedly concatenated on top of itself -- and correcting lower levels more frequently than higher levels -- a constant amount of error is tolerable, again with only polylogarithmic overhead.  The tolerable noise rate, which Aharonov and Ben-Or proved to be positive \cite{AharonovBenOr99}, is known as the fault-tolerance threshold.
Intuitively, small, constant-sized codes can be more efficient to use because encoding into the quantum code (which is necessary at the beginning of the computation and also during error correction, in certain schemes) is a threshold bottleneck.  
However, the threshold proof of Aharonov and Ben-Or only applies for concatenating codes of distance five or higher.  
In this paper, we prove a constant noise threshold for the concatenated distance-three, seven-qubit Steane/Hamming code.  (A threshold for the distance-three five-qubit code follows by the same structure of arguments.)  

The attainable threshold value, and the overhead required to attain it, are together of considerable experimental interest.  
Thus, while work has continued on proving the existence of constant thresholds in different settings -- e.g., under physical locality constraints \cite{Gottesman00local}, or with non-Markovian noise \cite{TerhalBurkard04} -- a substantial amount of attention has been devoted to estimating the fault-tolerance threshold, using simulations and analytic modeling.  
Most of these threshold estimates have used the seven-qubit code, 
from basic estimates \cite{Gottesman97thesis,Preskill97,KnillLaflammeZurekScience98,AharonovBenOr99}, to estimates using optimized fault-tolerance schemes \cite{Zalka97,Reichardt04,SvoreCrossChuangAho}, to a threshold estimate with a two-dimensional locality constraint \cite{SvoreTerhalDiVincenzo04}.  One reason the seven-qubit code has been so popular is no doubt its elegant simplicity, and its small size allows for easy, efficient simulations.  
However, there had been no proof that a threshold even existed for the simulated fault-tolerance schemes.  

Currently, the highest error threshold estimate is due to Knill, who has estimated a threshold perhaps as high as 5\% by using a very efficient distance-two code with a novel fault-tolerance scheme \cite{Knill05}.  Being of distance two, the code only allows for error detection, not correction, so the scheme uses extensive rejection testing.  This leads to an enormous overhead at high error rates, limiting the practicality of operating a quantum computer in this regime.  Still, a major open problem remains to prove the existence of a threshold for a distance-two code.

We prove an error threshold lower bound of $6.75 \times 10^{-6}$ in a certain error model.  Our analysis is prioritized for proof simplicity and ease of presentation, not for a high threshold (although we discuss optimizations in Sec.~\ref{s:estimates}).   
Also, 
it is not surprising that unproven threshold estimates should be significantly higher than proven threshold lower bounds -- although actually the author is unaware of any published rigorous lower bounds besides the current work and Ref.~\cite{AliferisGottesmanPreskill05} (except in the erasure error model \cite{Knill03erasure}).  
But such a large gap between what we can prove, and what our models and simulations indicate is embarrassing.  
A second major open problem is to close the gap between proofs and estimates.

Our proof is based on giving a recursive characterization of the probability distribution of errors in blocks of the concatenated code.  
Intuitively, with a distance-three code, two errors in a code block is a bad event, so the block error rate should drop roughly like $c p^2$, with $p$ the bit error rate.  After two levels of concatenation, the error rate should be like $c (cp^2)^2 = (cp)^{2^2}/c$, and so on\footnote{With $k$ levels of concatenation, the overall logical error rate is therefore $(c p)^{2^k}/c$.  To successfully implement an $N$-gate ideal circuit, we need an effective per-gate logical error rate $< 1/N$, requiring $k = O(\log \log N)$ concatenation levels.  With an $n$-qubit code, the encoding overhead is $n^k = \text{poly}(\log N)$.}.  The threshold for improvement is at $p=1/c$.  The difficulty lies in formalizing and making rigorous this intuition.  

The classic threshold proof of Aharonov and Ben-Or \cite{AharonovBenOr99} can be reformulated to rely on a key definition of 1-goodness.  Roughly, define a code block to be 1-good if it has at most one subblock which is not itself 1-good.  
Maintaining this definition as an inductive invariant through the logical circuit -- i.e., proving that the outputs of a logical gate are 1-good if the inputs are, with high probability -- 
allows provable thresholds for concatenating codes of distance five or higher.  
But this definition 
does not suffice for concatenating a distance-three code.  
For take a 1-good block, with the allowed one erroneous subblock, and apply a logical gate to it.  If a single subblock failure occurs while applying the logical gate, there can be two bad subblocks total, enough to flip the state of the whole block (since the distance is only three).  Therefore, the block failure rate is first-order in the subblock failure rate.  Logical behavior is not necessarily improved by encoding, and the basic premise of fault tolerance, controlling errors even with imperfect controls, is violated.  

Essentially, a stronger inductive assumption is required for the proof to go through. 
With 1-goodness, we are assuming that the block entering a computation step has no more than one bad subblock.  Intuitively, though, most of the time there should be no bad subblocks at all.  We capture this intuition in the stronger definition of ``1-wellness."   In a 1-well block, not only is there at most one bad subblock, but also the \emph{probability} of a bad subblock is small.  With this definition, the problem sketched above does not occur because the probability of there being an erroneous subblock in the input is already first-order, so a logical failure is still a second-order event.  
For the argument to go through, though, the definition must be carefully stated, and we need to carefully define what is required by each logical gate and how the logical gates will be implemented.  
Controlling the probability distribution of errors is the main technical tool and new contribution of this paper.

Very recently, Aliferis, Gottesman, and Preskill independently completed a threshold proof for distance-three codes \cite{AliferisGottesmanPreskill05}, based instead on formalizing the ``overlapping steps" threshold argument of Knill, Laflamme and Zurek \cite{KnillLaflammeZurekScience98}.  Our probabilistic definitions may be more difficult to extend to different error models.  
However, the probabilistic structure is also a potential strength, in that it may make this proof more extensible toward provable thresholds for postselection-based, error-detection fault-tolerance schemes like that of Knill \cite{Knill05}.

The intuition that small codes perform better needs to be qualified.
Small codes perform well because they can be quickly encoded and so allow frequent, rapid error correction at the lowest level.  However, larger codes offer protection with higher distances.    
There is therefore potentially a tradeoff.  
Steane finds that at noise rates well below the threshold it is best to start with a small code at the lowest concatenation level, then switch to a larger code \cite{Steane03}.  Also, if measurements are slow, then small codes can no longer perform rapid error correction, because they are waiting for measurements to complete.  In this case, Steane estimates that, from among a large set of codes, the seven-qubit code comes in only third behind the 23-qubit Golay code (distance 7) and a 47-qubit quadratic residue code (distance 11), which offer better efficiency compromises.  

Sections~\ref{s:definitions} and~\ref{s:stabilizer} contain the necessary definitions, and the proof of a threshold for quantum stabilizer operations (meaning preparation of fresh qubits as $\ket{0}$, measurement in the computational basis, and application of Clifford group unitaries like the CNOT gate).  
Stabilizer operations are easy to work with because Pauli errors (bit flips and dual phase flips) propagate through linearly.
While encoding and error-correction only require stabilizer operations, stabilizer operations do not form a universal gate set -- in fact, a circuit consisting only of quantum stabilizer operations can be efficiently classically simulated \cite{NielsenChuang00}.
Section~\ref{s:universality} extends the proof to give a threshold for full universal quantum computation, using the technique of ``magic states distillation" \cite{BravyiKitaev04,Reichardt04magic,Knill05}.  In fact, the threshold itself is unaffected by this extension -- the bottleneck in our threshold proof, as in most threshold estimates, is in achieving stabilizer operation fault-tolerance.

\section{Definitions} \label{s:definitions}

\subsection{Concatenated Steane code} \label{s:steanecode}

In concatenated coding,
qubits are arranged into 
level-one blocks of $n$, which are in turn arranged into level-two
blocks of $n$, and so on.  We call a single qubit a block$_0$, $n$ grouped qubits a block$_1$, and $n^k$ grouped qubits a block$_k$ (but often extraneous subscripts will be omitted).  

\begin{figure}  
\begin{center}
\begin{equation*}
\mspace{30mu}  
\left. \left.\begin{array}{c}
\Qcircuit @C=1.1em @R=.85em {
\lstick{\ket{+}}  & \ctrl{6}  & \qw & \qw \text{\LARGE $\mathcal{C}$} & \qw & & & & & &
\lstick{\ket{+}}  & \qw & \multigate{6}{\text{\LARGE $\mathcal{C}$}} & \qw \\
\lstick{\ket{+}}  & \qw & \ctrl{5} & \qw & \qw  & &  & & & &
\lstick{\ket{+}}  & \qw & \ghost{\text{\LARGE $\mathcal{C}$}} & \qw \\
\lstick{\ket{0}}  & \targ & \targ & \qw & \qw  & &  & & & &
\lstick{\ket{0}}  & \targ & \ghost{\text{\LARGE $\mathcal{C}$}} & \qw\\
\lstick{\ket{+}}  & \qw &  \qw  & \ctrl{3} & \qw  & &  & & & &
\lstick{\ket{+}}  & \qw & \ghost{\text{\LARGE $\mathcal{C}$}} & \qw\\
\lstick{\ket{0}}  & \targ &  \qw & \targ & \qw   & &  & & & &
\lstick{\ket{0}}  & \targ & \ghost{\text{\LARGE $\mathcal{C}$}} & \qw\\
\lstick{\ket{0}}  & \qw &  \targ & \targ & \qw  & &  & & & &
\lstick{\ket{\psi}}& \ctrl{-3} & \ghost{\text{\LARGE $\mathcal{C}$}} & \qw\\
\lstick{\ket{0}}  & \targ &  \targ & \targ & \qw \gategroup{1}{2}{7}{4}{.7em}{--}  & &  & & & &
\lstick{\ket{0}}  & \qw & \ghost{\text{\LARGE $\mathcal{C}$}} & \qw 
}
\end{array} \mspace{-236mu} \right\} \ket{0}_1 \mspace{183mu} \right\} \ket{\psi}_1
\end{equation*}
\end{center}
\caption{Circuit $\mathcal{C}$ (left) encodes the ancilla $\ket{0}_1$ using nine CNOT gates.  To encode an arbitrary state $\ket{\psi}_1$ requires $\mathcal{C}$ and two more CNOT gates (right).}\label{f:encoding}
\end{figure}
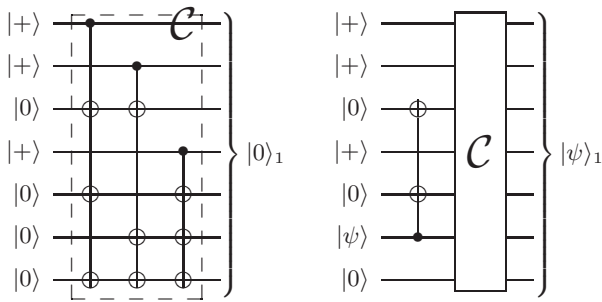

We will use the distance-three Steane code on $n=7$ qubits.  Recall its stabilizer generators
$$
\text{
\begin{tabular}{
c @{\!\,} c @{\!\,} c @{\!\,} c @{\!\,} c @{\!\,} c @{\!\,} c
@{,\,}
c @{\!\,} c @{\!\,} c @{\!\,} c @{\!\,} c @{\!\,} c @{\!\,} c
@{,\,}
c @{\!\,} c @{\!\,} c @{\!\,} c @{\!\,} c @{\!\,} c @{\!\,} c
}
I&I&I&Z&Z&Z&Z&
I&Z&Z&I&I&Z&Z&
Z&I&Z&I&Z&I&Z, \\
I&I&I&X&X&X&X&
I&X&X&I&I&X&X&
X&I&X&I&X&I&X,
\end{tabular}
}
$$
Here, $X = \left(\begin{smallmatrix}0&1\\1&0\end{smallmatrix}\right)$ is a bit flip operator -- $X \ket{0} = \ket{1}, X \ket{1} = \ket{0}$ -- while $Z = \left(\begin{smallmatrix}1&0\\0&-1\end{smallmatrix}\right)$ is a phase flip operator.  The first three stabilizer generators are equivalent to the classical $[7,4,3]$ Hamming code's parity checks, while the last three stabilizers are the same but in a dual basis.   (As is customary, tensor signs are omitted for legibility.)  
With this code, encoded, or logical, X and Z operators are transverse X and Z operators, respectively: i.e., $X_L = X^{\otimes 7}, Z_L = Z^{\otimes 7}$.
Other logical Clifford group operations are also performed by transverse physical Clifford group operations.  For example, the CNOT gate is defined by $\textsc{CNOT} \ket{a,b} = \ket{a,a \oplus b}$ for $a,b \in \{0,1\}$.  
Logical CNOT is just transverse CNOT.
Steane code encoding circuits are given in Fig.~\ref{f:encoding}.

\subsection{Error model}

\begin{definition}[Base error model] \label{t:baseerrors}
Assume each CNOT gate fails with probability $p$, independently of the others and earlier or simultaneous measurement outcomes, resulting in one of the sixteen Pauli products $I \otimes I, I \otimes X, \ldots, Z \otimes Z$ being applied to the involved qubits after a perfect CNOT  gate.
Assume that single-qubit Clifford group operations are perfect, that single-qubit preparation and measurement is perfect, and that there is no memory error.
\end{definition}

For ease of exposition, we consider a very simple error model, 
defined to start only for stabilizer operations.
Several of these assumptions are not essential.  There being no memory error is essential only as it implies arbitrary control parallelism, which is essential in threshold schemes.  The independence assumption of the CNOT failures can be relaxed as long as the as the conditional probability of failure (regardless of 
earlier or simultaneous
events) remains at most $p$.  The probabilistic nature of the failures is however essential for the proof in its current form.  Probabilistic failures of other operations can be straightforwardly incorporated.  

As is standard, 
additionally assume perfect classical control with feedback based on measurements.  All the required classical computations are efficient.  It is often assumed that classical computations are instantaneous, although this assumption doesn't matter with no memory error and even in general it can carefully be removed.  The use of feedback is important, particularly for the scheme we use to gain universality.  

\subsection{Error states}

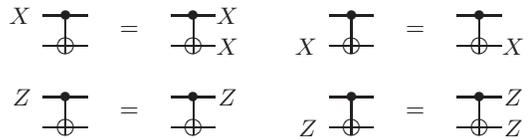
\begin{figure}
\begin{equation*}
\begin{array}{ccc}
\begin{array}{c}
\Qcircuit @C=.5em @R=.85em {
\lstick{X} & \ctrl{1} & \qw \\
               & \targ   & \qw 
}
\end{array}
\mspace{13mu}
=
\mspace{13mu}
\begin{array}{c}
\Qcircuit @C=.5em @R=.85em {
& \ctrl{1} & \qw & X \\
& \targ   & \qw & X
}
\end{array}
\mspace{40mu}
& &
\begin{array}{c}
\Qcircuit @C=.5em @R=.85em {
               & \ctrl{1} & \qw \\
\lstick{X}  & \targ   & \qw 
}
\end{array}
\mspace{13mu}
=
\mspace{13mu}
\begin{array}{c}
\Qcircuit @C=.5em @R=.85em {
& \ctrl{1} & \qw & \\
& \targ   & \qw & X
}
\end{array} 
\\
& & \\ 
\begin{array}{c}
\Qcircuit @C=.5em @R=.85em {
\lstick{Z} & \ctrl{1} & \qw \\
               & \targ   & \qw 
}
\end{array}
\mspace{13mu}
=
\mspace{13mu}
\begin{array}{c}
\Qcircuit @C=.5em @R=.85em {
& \ctrl{1} & \qw & Z \\
& \targ   & \qw & 
}
\end{array}
\mspace{40mu}
& &
\begin{array}{c}
\Qcircuit @C=.5em @R=.85em {
               & \ctrl{1} & \qw \\
\lstick{Z}  & \targ   & \qw 
}
\end{array}
\mspace{13mu}
=
\mspace{13mu}
\begin{array}{c}\Qcircuit @C=.5em @R=.85em {
& \ctrl{1} & \qw & Z \\
& \targ   & \qw & Z
}\end{array}
\end{array}
\end{equation*}
\caption{Propagation of Pauli errors through a CNOT gate; X errors are copied forward and Z errors copied backward.} \label{f:propagation}
\end{figure}

X, Z or Y $= i X Z$ errors are tracked from their introduction through the circuit by the commutation rules of Fig.~\ref{f:propagation}.  Some errors may have trivial effect -- e.g., $Z \ket{0} = \ket{0}$ -- but we still record them; we do not reduce errors modulo the stabilizer.  When we later extend the proof to a universal gate set, top-level logical Pauli errors can no longer be traced through the circuit, but we will make certain that these errors happen with vanishing probability.  

Since the code has distance three, 
error-free decoding of a block$_1$ is well-defined.  Error-free (perfect), bottom-up decoding of a block$_k$ is defined recursively by first decoding its subblocks$_{k-1}$, to interpret their states$_{k-1}$, then decoding the block.  Note that this recursive procedure is not the same as correcting to the closest codeword, but it is easier to analyze.

\begin{definition}[State]
The state$_0$ of a qubit is either I, X, Y, or Z, depending on what we have tracked onto that bit.  The state$_k$ of a block$_k$ is I, X, Y, or Z, determined by error-free decoding of the states$_{k-1}$ of its subblocks.
\end{definition}

The state of a block is determined by the states of its subblocks.  We want to define the \emph{relative} states of the subblocks, 
because a probability distribution over subblock errors is most naturally defined keeping in mind (i.e., relative to) the state of the enclosing block.  If a block is in error, then necessarily some of its subblocks will be in error.  

As a simple example, consider the classical three-bit repetition code: $0_L = 000, 1_L = 111$.  
If the states of three bits are XII -- the first bit is in error -- then the block's state decodes to be I.  The first bit is also in relative error.  
If the states of three bits are IXX, then the block's state decodes to be X.  The first bit is said to be in relative error (although it is not in error).
Making this definition precise, particularly in the quantum case, requires some care because different errors can be equivalent.

\begin{definition}[Relative syndrome$_k$]
The relative syndrome$_k$ of a block$_k$ consist of the syndromes of the $n-1$ code stabilizer generators on the states$_{k-1}$ of the subblocks$_{k-1}$.
\end{definition}

\begin{definition}[Relative state$_{k-1}$]\label{t:relativestate}
The relative states$_{k-1}$ of subblocks$_{k-1}$ of a block$_k$ are given by the minimum weight error, counting Y errors as two, generating the relative syndrome$_k$ of the block$_k$.  
\end{definition}
There is a unique minimum weight error, so this notion 
is well-defined, since every error syndrome can be achieved with at most one X and one Z error (or with one Y error).  
Note that according to this definition, relative states$_{k-1}$ are intuitively ``relative" to the state$_k$, and are not recursively related to the relative state$_k$.  

For example if the first two subblocks are in X error, then the block's state is X error, with the third subblock in relative X error (since logical X is equivalent to XXXIIII).  
Here is another example
with a state$_k$ of X:
$$\text{
\begin{tabular}{r @{\quad} c @{\!\;} c @{\!\;} c @{\!\;} c @{\!\;} c @{\!\;} c @{\!\;} c @{\!\;} l}
\text{states$_{k-1}:$}              & I&I&I&I&I&Y&X &\\
\text{relative states$_{k-1}:$} & X&I&I&I&I&Z&I &.
\end{tabular}
}$$

Unlike its state, the (X or Z component of the) relative state of a subblock can be determined by measuring the block transversely (in the Z or X eigenbases).  Again using the three-qubit repetition code for an example, $\ket{+}_L = \tfrac{1}{\sqrt{2}} (\ket{000} + \ket{111})$.   Since $(XII) \ket{+}_L = (IXX) \ket{+}_L \propto \ket{100} + \ket{011}$, one can't measure the state of the first qubit.  However, measuring in the 0/1 computational basis (Z eigenbasis) gives 100 or 011 with equal probabilities, telling us in either case that qubit 1 was in relative error (before the destructive measurement).

\subsection{Logical error model}

\begin{definition}[Logical error model] \label{t:logicalerrors}
The implementation of a logical operation $U_k$ on one or more blocks$_k$ is said to have had the correct logical effect if the following diagram commutes:
$$
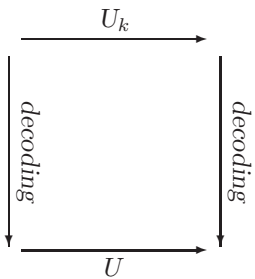
\begin{diagram}[balance,width=4em,height=4em,tight]
\, & \rTo{\text{U$_k$}}  & \, \\
\dTo>{\rotatebox{-90}{\makebox[0pt]{decoding}}} &  & \dTo>{\rotatebox{-90}{\makebox[0pt]{decoding}}} \\
\, & \rTo_{U} & \,
\end{diagram}
$$
Here the vertical arrows indicate perfect recursive decoding of the involved blocks, and the lower horizontal arrow represents a perfect $U$ on the decoded blocks.  

$U_k$ has had an incorrect logical effect if the same diagram commutes but with  $P \circ U$ on the bottom arrow, where $P$ is a Pauli operator or Pauli product on the involved blocks.
\end{definition}
In our error model, with our implementations, every logical operation will have either the correct logical effect or an incorrect logical effect
with some $P$ probabilistically.
For example, in error correction of a block, $U$ is the identity.  Error correction has the correct logical effect (no logical effect) if the state of the system following a perfect recursive decoding on the corrected blocks is the same as if we had just perfectly decoded the input blocks (and then applied the identity).  In particular, this implies that the state$_k$ of the output is the same as the state$_k$ of the input, but, more than that, also logical entanglement is preserved.  

\subsection{Goodness and wellness}

A key problem in proving a threshold is in establishing the proper definitions for inductively controlling the errors.  Once the correct definitions have been stated carefully, proving the relationships among them needed for a threshold result is fairly straightforward.  

The classic proof of a threshold in this setting, due to Aharonov and Ben-Or \cite{AharonovBenOr99}, can be framed as relying on the definition:

\begin{definition}[$r$-good$_k$] \label{t:good}
A block$_k$ is $r$-good$_k$ (and not $r\text{-bad}_k$) if it has $\leq r$ subblocks$_{k-1}$ which are either in relative error or not $r$-good$_{k-1}$.  A block$_0$ (single qubit) is $r$-good$_0$ if it is not in relative error.  
\end{definition}

So in a 1-good block, we have control over errors in $n-1$ of the subblocks (they are 1-good themselves and not in relative error), but potentially no control over the state of one of the subblocks.  

Definition~\ref{t:good} does not suffice for proving a threshold for a distance-three code because there is no room for errors in blocks which interact.  We can't maintain the inductive assumption of each block being 1-good because as soon as two blocks interact, they will then each have two subblocks with uncontrolled errors (with a priori constant probability, not second-order probability as we desire).  Aliferis et al. manage to use a similar definition, but change the method of induction proof to involve ``overlapping steps."  We instead will use a similar inductive proof to Aharonov and Ben-Or, but with different definitions
to look at probability distributions of errors.  

To give a threshold argument with a $d=3$ code, we will use a definition for probability distributions over relative errors.  

\begin{definition}[well$_k$]\label{t:well} 
A block$_k$ is well$_k$($p_1,\ldots,p_k$) if, conditioned on its state and on the errors in all other blocks, it has at most one subblock either in relative error or not well$_{k-1}(p_1,\ldots,p_{k-1})$, and
$$
\bf{P}(\text{such an uncontrolled subblock}) \leq p_k .
$$
\end{definition}

This definition conditions on the errors in all other blocks as a measure of independence.  Note that we don't assert anything about the distribution of errors within a subblock in relative error.  This is important because errors within relatively erroneous subblocks are typically less well controlled.  (For example, consider a perfect 7-qubit block$_1$, and introduce bitwise independent errors.  When the block as a whole is in error, one relative bit error is more likely than none, since two bit errors are more likely than three.)

For example, again using the three-qubit repetition code, the ensemble 
\begin{eqnarray*}
III & \text{w/ prob. $1-p$} \\
IXX & \text{w/ prob. $p$}
\end{eqnarray*}
is \emph{not} well$_1(p)$.  Even though the probability of a relative error is $\leq p$, conditioning on a logical state of X there is a relative error with probability one.  

Definition~\ref{t:well} can be generalized to $r$-well$_k$, a requirement on probability distributions with up to $r$ relative errors, but for a concatenated distance-three code threshold, $r=1$ is sufficient and so I have omitted any prefix.

\section{Fault-tolerance for stabilizer operations} \label{s:stabilizer}

\begin{figure*}
\begin{center}
\begin{equation*}
\begin{array}{c}
\Qcircuit @C=.35cm @R=.3cm {
& \ctrl{5} & \qw \\
& & \\
& & \\
& & \\
& & \\
& \targ & \qw 
}
\end{array}
\mspace{-28mu}
\raisebox{-26pt}{{$ _k$}}
\mspace{28mu}
\mspace{20mu}
\equiv
\mspace{30mu}
\begin{array}{c}
\Qcircuit @C=.35cm @R=.3cm {
& \ctrl{5} & \qw & \qw & \qw & \multigate{2}{EC_k} & \qw \\
\vdots & & \ddots & & & & \vdots \\
& \qw    & \qw & \ctrl{5} & \qw & \ghost{EC_k} & \qw\\
& & & & & & \\
& & & & & & \\
& \targ & \qw & \qw & \qw & \multigate{2}{EC_k} & \qw \\
\vdots & & \ddots & & & & \vdots \\
& \qw & \qw & \targ & \qw & \ghost{EC_k} & \qw
}
\end{array}
\mspace{-175mu}
\raisebox{-18pt}{\tiny{$ _{k-1}$}}
\mspace{175mu}
\mspace{-147mu}
\raisebox{-43pt}{\tiny{$ _{k-1}$}}
\mspace{147mu}
\end{equation*}
\end{center}
\begin{equation*}
\begin{array}{c}
\Qcircuit @C=.35cm @R=.3cm {
& \gate{EC_k} & \qw
}
\end{array}
\mspace{20mu}
\equiv
\mspace{30mu}
\begin{array}{c}
\Qcircuit @C=.25cm @R=.3cm {
 & \gate{EC_{k-1}} & \ctrl{1} & \targ &   \qw&\qw&\qw&\qw&\qw&\qw  & \gate{EC_{k-1}} & \ctrl{1} & \targ & \qw & \qw &\qw & \qw & \qw \\
 & \lstick{\ket{+}_k} & \targ & \qw & \meter & & & & & & \lstick{\ket{+}_k} & \targ & \qw & \meter & & & & \\
 & \lstick{\ket{0}_k} & \qw & \ctrl{-2} & \meter & & & & & & \lstick{\ket{0}_k} & \qw & \ctrl{-2} & \meter & & & &
 }
\end{array}
\def\place #1#2#3{\mspace{#2}\makebox[0pt]{\raisebox{#3}{#1}}\mspace{-#2}}
\place{\tiny{$ _{k-1}$}}{-378mu}{-2pt}
\place{\tiny{$ _{k-1}$}}{-360mu}{-23pt}
\place{\tiny{0/1}}{-300mu}{0pt}
\place{\tiny{+/-}}{-300mu}{-21pt}
\place{\tiny{$ _{k-1}$}}{-128mu}{-2pt}
\place{\tiny{$ _{k-1}$}}{-110mu}{-23pt}
\place{\tiny{0/1}}{-50mu}{0pt}
\place{\tiny{+/-}}{-50mu}{-21pt}
\place{\rotatebox{70}{\mbox{\tiny{apply correction $\rightarrow$}}}}{-270mu}{-32pt}
\place{\rotatebox{70}{\mbox{\tiny{apply correction $\rightarrow$}}}}{-20mu}{-32pt}
\place{/}{-479mu}{27pt}
\place{/}{-229mu}{27pt}
\place{/}{-394mu}{10pt}
\place{/}{-144mu}{10pt}
\place{/}{-365mu}{10pt}
\place{/}{-115mu}{10pt}
\end{equation*}
\caption{A concatenation-level-$k$ encoded CNOT, or CNOT$_k$, acting on two blocks each with $n$ subblocks ($n^k$ qubits in all), is implemented as transverse CNOTs$_{k-1}$ (i.e., CNOTs$_{k-1}$ from subblock $i$ of the first input block to subblock $i$ of the second input block, $i=1,\ldots, n$) followed by level-$k$ error corrections on each block.  Error correction$_k$ is implemented as transverse error corrections$_{k-1}$, followed by transverse CNOTs$_{k-1}$ to and from prepared ancillas$_k$.  These ancillas are measured, and an appropriate correction (a level-$(k-1)$ encoded operation) is possibly applied.  Then the whole process is repeated.}
\label{f:implementations}
\end{figure*}
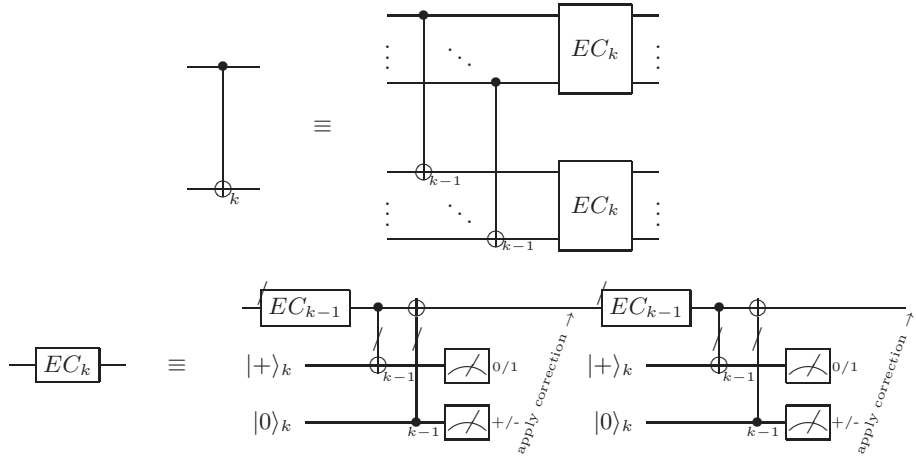

Our proof of fault-tolerance for stabilizer operations will rely on three indexed Claims A$_k$, B$_k$ and C$_k$ for, respectively, encoded ancilla preparation, error correction and encoded CNOT, at code-concatenation level $k$, with the following inductive dependencies:
$$
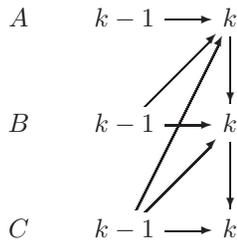
\begin{diagram}[balance,width=2em,height=2em,tight]
A & & k-1 & \rTo     & k \\
   & &       &  \ruTo \ruTo(2,4)         & \dTo  \\
B & & k-1 & \rTo          & k \\
   & &       & \ruTo     & \dTo  \\
C & & k-1 & \rTo     & k 
\end{diagram}
$$
That is, as shown in Fig.~\ref{f:implementations}, a level-$k$ encoded CNOT, or CNOT$_k$, will use CNOTs$_{k-1}$ and error corrections$_k$ -- so the proof of Claim C$_k$ will rely on Claims C$_{k-1}$ and B$_k$.  Error correction$_k$ will use corrections$_{k-1}$ and CNOTs$_{k-1}$, as well as ancillas$_k$.  Finally, the proof of Claim A$_k$ (ancilla$_k$ preparation) will rely on each of Claims A$_{k-1}$, B$_{k-1}$ and C$_{k-1}$.

Each level-$k$ operation will fail with probability $A_k$, $B_k$ or $C_k$ (failure parameters are italicized unlike the names of the claims to which they correspond).  Failure parameters will drop quadratically at each level, giving a threshold as sketched in Sect.~\ref{s:introduction}.  That is, 
$$
\max \{A_k, B_k, C_k\} = O\!\left((\max \{A_{k-1}, B_{k-1}, C_{k-1}\})^2\right) .
$$
Splitting out separate error parameters in this way lets us easily track where errors are coming from, and lets us find the threshold bottlenecks for optimization.
We will also define two wellness parameters $a_k$ and $b_k$ (since a CNOT$_k$ ends with error corrections$_k$, there is no need for a separate wellness parameter $c_k$).

Relaxing some of the assumptions in our error model would just require modifying these claims, and possibly adding new ones.  For example, we have assumed perfect measurements, but faulty measurements would only require a fourth indexed claim dependent only on itself 
(a level-$k$ measurement outcome is the decoding of $n$ level-$(k-1)$ measurement outcomes).  

Preparation of a single-qubit ancilla in state $\ket{0}_0 = \ket{0}$ or $\ket{+}_0 = \ket{+} \equiv \tfrac{1}{\sqrt{2}}(\ket{0}+\ket{1})$ we assume to be perfect, so define $A_0 \equiv 0$.  (Alternatively, set $A_0 > 0$ to remove this assumption.)
For $k \geq 1$, we need:

\theoremstyle{theorem}
\newtheorem*{claimA}{Claim A$_k$}
\begin{claimA}[Ancilla$_k$] \label{t:claimA}
Except with failure probability at most $A_k$, we can prepare a level $k$ ancilla $\ket{0}_k$ which has a state$_k$ of I (no error) and is well$_k$($b_1,\ldots, b_{k-1}, a_k$).
\end{claimA}
\noindent (The different parameters will be set explicitly within the proofs.)

Error correction$_k$ is only defined for $k \geq 1$ levels of encoding, so we may take $B_0 \equiv 0$.

\theoremstyle{theorem}
\newtheorem*{claimB}{Claim B$_k$}
\begin{claimB}[Correction$_k$] \label{t:claimB}
With probability at least $1-B_k$, the output is well$_k$($b_1, \ldots, b_k$) and, if the input is well$_k$($b_1, \ldots, b_k$), there is no logical effect. 

Additionally, if all but one of the input subblocks$_{k-1}$ are well$_{k-1}$($b_1,\ldots,b_{k-1}$) and not in relative error, then with probability at least $1-B_k'$ there is no logical effect and the output is well$_k$($b_1,\ldots,b_k$).
\end{claimB}

Note how powerful a successful correction$_k$ is.  Even if there is no control whatsoever on the errors in the input block, the output is still well$_k$.  This property is essential for getting errors fixed in a fully recursive manner, because it means that to fix an erroneous subblock$_{k-1}$, we need only apply a single-qubit correction transversely on that subblock and there is no need to worry about bit errors within that subblock.

CNOT$_0$ is simply a physical CNOT gate.  $C_0 \equiv p$ is the probability of failure of a physical CNOT gate (for the oft-used simultaneous depolarization error model, $p$ is $\tfrac{15}{16}$ times the depolarization rate).  

\theoremstyle{theorem}
\newtheorem*{claimC}{Claim C$_k$}
\begin{claimC}[CNOT$_k$] \label{t:claimC}
With probability at least $1-C_k$, the output blocks$_k$ are well$_k$($b_1, \ldots, b_k$) and, if the input blocks$_k$ are well$_k$($b_1, \ldots, b_k$), then a logical CNOT, the correct logical effect, is applied. 
\end{claimC}

\begin{proof}[Proof of Claim C$_k$]
CNOT$_k$ is implemented as transverse CNOT$_{k-1}$, followed by Correction$_k$ on each block.

Always declare failure if either Correction$_k$ fails.  This gives the desired behavior on the case of general inputs with $C_k \geq 2 B_k$; on success, the outputs are well$_k$ because they are leaving successful Corrections$_k$.  Assume now that the input blocks are each well$_k$($b_1,\ldots,b_k$).

If none of the input subblocks are in relative error (probability at least $1-2b_k$), we say the CNOT$_k$ has failed if either of the Corrections$_k$ fail, or if two or more CNOTs$_{k-1}$ fail.  The conditional failure rate is 
$$
\leq (1-C_{k-1})^n 2 B_k + n C_{k-1} 2B_k' + \binomial{n}{2} C_{k-1}^2 .
$$

If one of the input subblocks is in relative error (prob. $\leq 2b_k$), we say the CNOT$_k$ has failed if either of the Corrections$_k$ fail, or if one or more CNOTs$_{k-1}$ fail.  The conditional failure rate is 
$$
\leq (1-C_{k-1})^n 2B_k' + n C_{k-1} .
$$

If two input subblocks are in relative error (prob. $\leq b_k^2$), we say the CNOT$_k$ has failed.  

On success, the correct logical effect has been applied (Def.~\ref{t:logicalerrors}).  Indeed, the successful corrections have no logical effect, so the ideal decoding can be commuted past them.  All but possibly one of the CNOTs$_{k-1}$ have the correct logical effect, so decoding up to level $k-1$ can be commuted past them, too, leaving ideal transverse CNOT gates.  Transverse CNOT gates are an encoded/logical CNOT gate for the Steane code (Sec.~\ref{s:steanecode}).  The final decoding round can tolerate one subblock in error from each block so it also commutes forward, leaving a single ideal CNOT gate.
Overall, the probability of failure is at most 
\begin{equation*}\begin{split}
C_k &\equiv 
(2 B_k + 2 n C_{k-1} B_k' + \binomial{n}{2} C_{k-1}^2) \\
&\quad + 2 b_k (2 B_k' + n C_{k-1}) + b_k^2 . \qedhere
\end{split}\end{equation*}  
\end{proof}

\begin{proof}[Proof of Claim B$_k$]
Correction$_k$, shown in Fig.~\ref{f:implementations}, begins with transverse Corrections$_{k-1}$.  Then correct for X relative errors using a Z syndrome extraction (explained below), and symmetrically correct for Z relative errors.  Then repeat:  Apply transverse Corrections$_{k-1}$ and correct for X and Z relative errors again.

To extract a Z syndrome, used for correcting X (bit flip) relative errors, apply transverse CNOTs$_{k-1}$ into a $\ket{+}_k$ ancilla \cite{Steane03}.  (Logically, there should be no effect, since $\ket{+} = \tfrac{1}{\sqrt{2}}(\ket{0} + \ket{1})$ is an eigenstate of the NOT/X operation.  But as in Fig.~\ref{f:propagation}, X errors will be copied into the ancilla.)  Measure this ancilla transversely in the 0/1 computational basis (Z eigenbasis) and bottom-up recursively decode in order to determine which ancilla subblock, if any, is in X relative error.  
Apply X gates transversely to this subblock as a correction\footnote{From Def.~\ref{t:baseerrors}, we are assuming perfect X and Z gates, and no memory error.  In fact, though, it isn't necessary to apply the correction as long as the experimenter classically tracks what the correction should have been through the rest of the computation.}.

We give only a combinatorially loose analysis; some optimizations are described in Sec.~\ref{s:estimates}.
There are three cases, depending on the input.

Always declare failure if any of the four ancilla preparations fails.  

1. If the input is already well$_k$($b_1,\ldots,b_k$), declare failure if there are two level $k-1$ failures.  Overall, there are $2n$ Corrections$_{k-1}$, 
$4n$ CNOTs$_{k-1}$, and four prepared ancillas$_k$, so the probability of failure is at most 
\begin{equation*}\begin{split}
B_k &\equiv  4 A_k + \binomial{4n}{2} C_{k-1}^2 + \binomial{4}{2} a_k^2 + \binomial{2n}{2}B_{k-1}^2 \\
&\quad + b_k (4n C_{k-1} + 4 a_k + 2n B_{k-1}) \\
&\quad + 4n C_{k-1} (4 a_k + 2n B_{k-1}) + 8n a_k B_{k-1} .
\end{split}\end{equation*}
We can conservatively set 
the block output wellness parameter
$b_k$ to be 
the probability of any single level $k-1$ failure occurring in the correction circuit -- not in the input:
$$
b_k \geq \tilde{b}_k \equiv (4 a_k + 2n B_{k-1} + 4n C_{k-1}) / (1-B_k) .
$$
Relative errors in the input will be corrected and won't show up as relative errors in the output (assuming success).  

By combining commutative diagrams from Def.~\ref{t:logicalerrors}, clearly on success the output is well and no logical operation has been applied.  

2. In the case that all but one of the input subblocks are well$_k$($b_1,\ldots,b_{k-1}$) and not in relative error, again the output will be well with probability at least $1-B_k$.  However, a single level $k-1$ failure can change the state of the block, since the code has only distance three.  We introduce the failure parameter $B_k' \equiv 4 A_k + (1-B_k) \tilde{b}_k$ to bound the probability that the state changes.  (This is too conservative since only a single one of the early level $k-1$ operations can change the state; later on, two level $k-1$ failures are still required.)  
Set $b_k \equiv (1-B_k) \tilde{b}_k / (1-B'_k)$.  In the parameter range of interest, $b_k \geq \tilde{b}_k$.

3. Consider now the most interesting case, which has dictated our implementation of Correction$_k$: when the input is uncontrolled.  
(In either of the two previous cases, 
one round of 
transverse Corrections$_{k-1}$ followed by X and Z correction 
would have sufficed.)  
In this case, we can make no guarantees about the logical effect of the correction, but want the output to be well.

Here is an example which shows why this arbitrary input case is more complicated: Say that exactly all input subblocks are well, with exactly one in error.  In this case, one level $k-1$ failure in the first X and Z correction rounds can change the whole block's state.  
This is okay; we are making no guarantees about the output state versus the input state.   
The problem is that conditioning on the block's state having changed (as in the definition of well), there is a zeroth-order probability of a subblock in relative error.  Thus the output after the first rounds of X and Z correction is not necessarily well.  

If there are no level $k-1$ failures in the first X and Z correction rounds, then the block halfway through will already have all its subblocks well and none in relative error.  The second X and Z correction rounds will leave the block well.  

If there is a level $k-1$ failure in the first X and Z correction rounds, then leaving these rounds there can be one subblocks which is not well and one subblock which is in relative error, as in the example.  However, no level $k-1$ failures are then allowed in the second X and Z correction rounds.  The transverse Corrections$_{k-1}$ will fix the subblock (if any) which is not well.  The X and Z correction rounds will fix the subblock in relative error.  
\end{proof}

Ancilla preparation is a key step in a fault-tolerance scheme.  (This is particularly true for schemes based on teleportation, like those of Knill \cite{Knill03erasure,Knill05}, and in our own method for achieving universality in Sect.~\ref{s:universality}.)  
Only during ancilla preparation do subblocks within a particular block interact with each other.  The preparation scheme must dampen these interactions to avoid strong correlations between errors in different subblocks.  The proof of Claim~A$_k$ runs along similar lines as the above proofs, and is given in Appendix~\ref{s:proofak}.  The failure parameters determined there are 
\begin{eqnarray*}
A_k 
&\equiv&
\left( \begin{split}
\binomial{2n}{2} (A_{k-1}^2 + B_{k-1}^2)  + \binomial{2s(n)+n}{2} C_{k-1}^2 \\
+ 2n(2s(n)+n) (A_{k-1} + B_{k-1}) C_{k-1} \\
+ 4n^2 A_{k-1}B_{k-1} 
\end{split}\right) / N_k  \\
a_k
&\equiv& 
\frac{ 2n (A_{k-1} + B_{k-1}) + (2s(n)+n) C_{k-1} }{ (1-A_k) N_k } ,
\end{eqnarray*}
where we may take $s(n=7)=9$ and 
$$
N_k \equiv 1- 2 n A_{k-1} - 2 n B_{k-1} - (2 s(n) + n)C_{k-1} .
$$

\section{Estimates of the constant threshold} \label{s:estimates}

The results in Sec.~\ref{s:stabilizer} give the claimed constant threshold for stabilizer operations, because the error parameters each drop quadratically at each level of concatenation.  This is straightforward to prove rigorously by replacing complex inequalities with simpler upper bounds.  

Our goal was to complete a rigorous proof of a constant threshold, not to estimate the true threshold.  Still, it is interesting how high a threshold these techniques give us.  We have numerically iterated the equations of Sec.~\ref{s:stabilizer}, taking $n=7$. 
We found that the error rates converged to zero for $p = C_0 < 6.75 \times 10^{-6}$.  Of course this is not a \emph{proof} that the equations converge 
in this range, 
but the proof is doable: simply bound arithmetic errors while iterating the equations numerically up to a level at which weaker bounds can be applied to show convergence. 

This threshold does not compare directly to the $4.18 \times 10^{-5}$ rigorous threshold lower bound of 
Aliferis et al. \cite{AliferisGottesmanPreskill05} because their error model allows faults in single-qubit preparation and measurement as well as just CNOT gates.  This means that preparing an unverified ancilla, for example, has 16 possible fault locations instead of just 9.  
We expect that incorporating these single-qubit errors into our model would reduce the threshold by a small constant.  
Also, our recursion equations were highly conservative, typically bounding the probability of a level $k$ failure by the probability of any two level $k-1$ failures.  Aliferis et al., however, used a computer to count exactly which pairs of level $k-1$ faults could cause a level $k$ failure.  This counting obviously directly improves their threshold estimate, but it improves the threshold estimate in an indirect way, too; it allows them to use more easily less modular constructions.

There are a number of optimizations that can be carried out to improve the threshold slightly.  
We have briefly investigated improved ancilla preparation, improved implementation of CNOT$_k$, and  improved analysis techniques. 

For example, in Claim A$_k$, we apply transverse Correction$_{k-1}$ at the end of preparing an unverified ancilla.  But each CNOT$_{k-1}$ ends with Corrections$_{k-1}$ anyway, so this is redundant and useful only for a more modular analysis.  
Ancilla preparation can also be improved by verifying against both X and Z errors.  

Actually, though, with CNOT$_{k-1}$ notice that if the input blocks are well, then full, two-round Correction$_{k-1}$ is not needed on each output block -- instead, a single round, using two ancillas instead of four, suffices.  In unverified ancilla preparation with this version of CNOT$_{k-1}$ a transverse, two-round Correction$_{k-1}$ \emph{would} be needed for the proof to go through, not just for modularity. 
Thus if we change our CNOT operations to be followed by a one-round correction, 
then during unverified ancilla preparation, seven full corrections and eighteen one-round corrections total will be applied (or, sacrificing some modularity, seven full and eleven one-round corrections), as compared to eighteen full corrections currently.  This increases the threshold 
to $1.46 \times 10^{-5}$, 
according to numerical iterations of the recursion equations.  

We might also split out two different kinds of CNOT$_k$ gates, because if one of the output blocks of a CNOT gate is to be measured immediately (as is the case for the transverse CNOT gates in preparation), then error correction of that block can only introduce more errors.  
Another optimization is to switch to error-correction via teleportation \cite{Knill03erasure}.
A factor of two or three improvement in the proven threshold seems to be in reach using these kinds of optimizations -- which are easy to write on the back of an envelope, but tedious to write out formally.

The bottlenecks in our equations in fact appear to be ancilla preparation subblock failure $a_k$ and CNOT failure C$_k$.  It should be productive to improve the analysis of ancilla preparation -- by for example verifying against Z as well as X errors, and by using computer counting to minimize the overhead from artificial modularity.  Also, during the counting, one should try to distinguish X and Z errors as much as possible -- this might be simplest with a more specialized base error model.  Once the error model is nailed down, one can actually simulate ancilla preparation at the lower few levels and determine higher estimates for ancilla failure rates just at these levels.  This allows for analyzing more complicated ancilla preparation schemes, including schemes based more heavily on postselection \cite{Reichardt04}.  Assuming (or proving) these estimates to be trustworthy can increase the threshold significantly, even though only a few levels (e.g., $A_1, a_1, A_2, a_2$) are directly modified.

These threshold estimates for stabilizer operations remain applicable for full quantum universality, discussed in Sec.~\ref{s:universality}.  The techniques we use to gain universality can tolerate higher error rates than those for stabilizer operations, so achieving universality is not a bottleneck.

\section{Fault-tolerant universality} \label{s:universality}

To achieve universality, we use the technique of ``magic states distillation" \cite{BravyiKitaev04,Reichardt04magic,Knill05}.  It is well known that stabilizer operations plus the ability to repeatedly prepare one of the ``magic," single-qubit pure states $\ket{H}$ or $\ket{T}$ gives quantum universality.  (More generally, in fact, stabilizer operations plus the ability to repeatedly prepare any single-qubit pure state which is not a Pauli eigenstate gives quantum universality.)  Magic states distillation is a technique for using perfect stabilizer operations to distill faulty copies of $\ket{H}$ or $\ket{T}$ to perfect copies (or at least, arbitrarily close to perfect with only polylogarithmic overhead).  

Operating beneath the threshold for stabilizer operations, we can assume the error rate is arbitrarily small, and therefore condition on no stabilizer operation errors at all.  So we have perfect stabilizer operations.  
To apply the technique of magic states distillation to achieve universality, following Knill \cite{Knill04schemes,Knill05}, we create an encoded Bell pair.  Then we decode one half from the bottom up.  (Unlike Knill, we do not postselect on no detected errors.)  

At this point, we have a Bell pair, half of which is a single qubit unprotected from errors, and the other half is encoded.  Teleport a single-qubit (approximate) ``magic state" $\ket{H}$ or $\ket{T}$ into the encoding, using a physical CNOT gate and two single-qubit measurements.  Of course, there can be errors in the teleportation.  But any errors will also be teleported, into logical errors on the remaining encoded $\ket{H}$ or $\ket{T}$.  Encoded stabilizer operations can then distill these into perfect logical magic states (meaning no logical errors, bit errors of course remain), from which we obtain encoded universality.  

To prove that this scheme works, we need to understand the behavior of the recursive, bottom-up decoding procedure.  It is fairly straightforward that this works, maintaining a bounded error rate; a simple proof is given below.  
We also need to verify that the magic states distillation procedures apply in our setting.

\subsection{Decoding}

\theoremstyle{theorem}
\newtheorem*{lemmaD1}{Lemma D$_1$}
\begin{lemmaD1}[Decoding$_1$] \label{t:decoding1}
Consider a well$_1$($p_1$) block$_1$.  On each bit independently, apply a Pauli error with probability at most $q_1$.  
Decoding of the block can still be done successfully -- meaning the state of the system is the same as if perfect decoding had been applied -- with probability at least $1-D_1(p_1,q_1)$.  The rest of the time, the system's state is the same as if either I, X, Y or Z had been applied after perfect decoding.
\end{lemmaD1}

\begin{proof}
If any CNOT gate fails, then an error can be introduced (there is no fault tolerance).  Additionally, an error will be introduced if there are two or more errors in the input after the bit errors are applied, which occurs with probability at most $n p_1 q_1 + \binomial{n}{2} q_1^2$.  
Letting $e(n)$ bound the number of CNOT gates required for encoding/decoding an $n$-qubit code ($e(n) = O(n^2)$ but is possibly $> s(n)$ defined in the proof of Claim~A$_k$ as the number of CNOTs required for encoding a stabilizer state), we may take
$$
D_1(p_1,q_1) \equiv e(n) C_0 + n p_1 q_1 + \binomial{n}{2} q_1^2 .
$$
For the Steane code, set $e(7) = 11$ (Fig.~\ref{f:encoding}).
\end{proof}

\theoremstyle{theorem}
\newtheorem*{claimDk}{Claim D$_k$}
\begin{claimDk}[Decoding$_k$] \label{t:decodingk}
Decoding of a well$_k$($b_1, \ldots, b_k$) block$_k$ can be done successfully -- meaning the state of the system is the same as if perfect decoding had been applied -- with probability at least $1-D_k$.  On failure, the system's state is the same as if either I, X, Y or Z is applied after perfect decoding.
\begin{gather*}
D_1 \equiv e(n) C_0 \\
D_k \equiv e(n) C_0 + n b_k D_{k-1} + \binomial{n}{2} D_{k-1}^2 .
\end{gather*}
\end{claimDk}

\begin{proof}
First apply Lemma~D$_1$ with $p_1 = b_1$ and $q_1 = 0$, in order to decode the bottom layer of a well$_2$($b_1, b_2$) block$_2$.  This outputs a well$_1$($b_2$) block$_1$, except each bit has an additional failure probability of at most $D_1$.  Thus the claim can be applied again with $p_1 = b_2$ and $q_1 = D_1$ to get a single bit out which will be correct except with a failure probability which we bound by $D_2$.  Continue inductively. 
\end{proof}

Clearly if $p = C_0$ is beneath the threshold for stabilizer operations (so $b_k$ converges to zero quadratically at each level), then too there is a threshold for $p$ beneath which $D_k$ converges to a constant.  In our numerical threshold estimates, the threshold for $D_k$ is equal to that for stabilizer operations.

\subsection{Magic states distillation}

Applying the magic states distillation results of Ref.'s \cite{BravyiKitaev04,Reichardt04magic} requires a little bit of work.  
The main concern here is that these results 
were derived under the assumption that multiple identical copies of $\rho$ the input faulty state can be prepared at will.
However, in our setting we can't assume that different encoded approximate magic states have exactly the same logical error rates.  
In fact, at least for the $\ket{T}$ state distillation scheme, the 
bound on the allowed error rate is the same for nonidentical states as for identical states.  (Numerical experiments indicate that the same is probably true for $\ket{H}$ state distillation schemes.)

The $x, y, z$ Bloch sphere coordinates of a single-qubit state $\rho$ are $\tfrac{1}{2}\tr (X.\rho), \tfrac{1}{2}\tr (Y.\rho), \tfrac{1}{2}\tr (Z.\rho)$, respectively.  Define $\ket{T}$ by coordinates $x=y=z = 1/\sqrt{3}$ of $\ketbra{T}{T}$.  Let $T = (e^{2\pi i /3} - 1) \ketbra{T}{T} + I$.  Before beginning distillation, symmetrize each state by applying $I$, $T$ or $T^2$ with equal probabilities $1/3$ -- so we may assume each state has equal $x, y, z$ coordinates.

\theoremstyle{theorem}
\newtheorem*{lemmamagic}{Lemma 1}
\begin{lemmamagic}
Perfect adaptive stabilizer operations together with the ability to prepare states $\rho_i$, $i = 1, 2, \ldots$, with Bloch sphere coordinates $x_i = y_i = z_i = f_i / \sqrt{3}$, such that $f_i \geq \sqrt{3/7} + \epsilon$, $\epsilon > 0$, gives quantum universality.  Here $\epsilon > 0$, a constant, is known, but the $f_i$ may not be.
\end{lemmamagic}

A proof is given in Appendix~\ref{s:moregeneralCSS}.  

\section{Conclusion}

We have proved:
\begin{theorem}
For the error model specified in Def.~\ref{t:baseerrors}, arbitrarily accurate, efficient, universal quantum computation is achievable, via a scheme based on concatenation of the $[\![7,1,3]\!]$ Steane code, as long as the error rate is beneath a positive constant threshold.
\end{theorem}

Note that repeated syndrome extraction during error correction is 
not necessary as long as the ancillas used in error corrections are prepared fault-tolerantly (we use a simple postselection scheme, but repeated syndrome extraction to verify ancillas would also have sufficed).  
It should also be emphasized that preparation of reliable encoded states is a major threshold bottleneck.  

A major open question in quantum fault-tolerance lies in rigorously proving higher thresholds.  As the current highest threshold estimates rely heavily on postselection, a good understanding of the probability distribution of errors seems to be necessary to prove results about these schemes.  Our arguments comprise a first, minor step in this direction, but to go further one needs to characterize the probability distribution of errors within the blocks which are in relative error.  One also needs to be concerned about any dependencies between blocks -- for even in very artificial models full independence is impossible to maintain (or even converge to asymptotically), and small deviations can rapidly build up into large dependencies when postselections discard large fractions of the probability mass.  Perhaps combining a probabilistic analysis with the overlapping steps method of Aliferis et al. could lessen the strict analysis requirements which make a rigorous proof in a postselected setting so difficult.  

There are many other open problems, of course.  
How high can the provable threshold be pushed for schemes not based on postselection?  We are still several orders of magnitude below some of the higher estimates.  One promising way to approach this problem is with a rigorous, but computer-aided, analysis of the lower levels of the fault-tolerance scheme, then ``plugging in" to a conservative, analytical estimate once the error rate has dropped sufficiently.  
Running Monte Carlo simulations, then fitting the simulation results to our failure and wellness parameters, is another method for understanding error behavior -- although then we would obtain only confidence intervals for the threshold.  
(Since the threshold bottleneck is in achieving fault-tolerant stabilizer operations, which are classically simulatable, simulations are efficient.)  

Can this threshold proof be extended to more general error models (not just probabilistic Pauli errors)?  
A real quantum computer is likely to use lower-level error-correction techniques highly specialized to the physical error model -- for example composite pulses \cite{ReichardtGrover05} -- before switching to a more general fault-tolerance scheme.  It is important to better understand the interface/transition between the low- and high-level error models/error-correction schemes.

Efficiency is a major practical concern;
even constants are very important.  
All constant-threshold fault-tolerance schemes should ultimately have the same efficiency in big-$O$ notation -- one can just switch to the most efficient scheme once a less efficient scheme has reduced the error rate sufficiently (with constant overhead).  
For practical quantum computing, a threshold result is not needed --
only an argument that the effective logical error rate, in some sense, can be pushed low enough to do interesting computations.  Dropping the requirement that the error model definition concatenates nicely might gain us some freedom.  
Steane has investigated concatenating on larger codes once a smaller code like the $[\![7,1,3]\!]$ code has reduced the error rate sufficiently, 
and found it to be an apparently useful technique \cite{Steane03}.

Is there a more efficient method of obtaining universality for our scheme?  Can one distill multi-qubit states which form a more natural universal gate set?  What can be done with a non-adaptive circuit structure? 

The present work merely sheds light on a fact which had long been assumed -- a threshold for a distance three code -- but not proved.  Hopefully a more solid foundation will help us as we try to address the more ambitious 
open questions in this field.

\begin{acknowledgments}
Research supported in part by NSF ITR Grant CCR-0121555, and ARO Grant DAAD 19-03-1-0082.
\end{acknowledgments}

\appendix 

\section{Proof of Claim~A$_k$} \label{s:proofak}

\begin{proof}[Proof of Claim A$_k$]
To prepare the ancilla$_k$, we first prepare $n=7$ ancillas$_{k-1}$ either $\ket{0}_{k-1}$ or $\ket{+}_{k-1}$, then apply $s(n)$ CNOTs$_{k-1}$ to entangle them, 
then apply Correction$_{k-1}$ transversely.  For the Steane code, $s(7) = 9$ (Fig.~\ref{f:encoding}) -- entangling a general $n$-qubit stabilizer state requires at most $s(n) = O(n^2)$ CNOTs.
(The ancilla $\ket{+}_{k-1}$ can be prepared by transversely applying [by assumption perfect] transverse Hadamard gates to $\ket{0}_{k-1}$, or can be prepared directly by symmetry.)  

Run the above procedure twice to create two unverified ancillas $\ket{0}_k$, and apply CNOT$_{k-1}$ transversely from the first copy to the second.  Then measure the second half in the Z eigenbasis (computational basis), postselecting on no X errors$_{k-1}$.  
That is, we measure all the physical qubits in the second half, then in a classical computer recursively decode up to the states$_{k-1}$.  If there are any relative errors$_{k-1}$ or if the state$_k$ is not I, then discard the ancilla qubits and start over.
(Discarding qubits based on measurement outcomes requires adaptive control.  Ancilla preparation can be implemented without adaptive control by verifying against two copies instead of one.   
However, our method of obtaining universality in Sec.~\ref{s:universality} will require adaptive control anyway.)

The probability of acceptance (normalization constant) is at least 
$$
N_k \equiv 1- 2 n A_{k-1} - 2 n B_{k-1} - (2 s(n) + n)C_{k-1} ,
$$
or one minus a unon bound on the probability of a level $k-1$ failure during the two unverified ancilla preparations and the transverse Corrections$_{k-1}$ and CNOTs$_{k-1}$.
In our parameter ranges, this will always be a constant, so the overhead from postselection is acceptable.  

We declare failure, conservatively, if there are two level $k-1$ failures and the ancilla is still accepted.  The probability of failure is at most 
$$
A_k \equiv \left( \begin{split}
\binomial{2n}{2} (A_{k-1}^2 + B_{k-1}^2)  + \binomial{2s(n)+n}{2} C_{k-1}^2 \\
+ 2n(2s(n)+n) (A_{k-1} + B_{k-1}) C_{k-1} \\
+ 4n^2 A_{k-1}B_{k-1} 
\end{split}\right) / N_k .
$$

If there are no level $k-1$ failures, 
then there are no errors$_{k-1}$ in the output ancilla, as can easily 
be verified by using the commutative diagram of Definition~\ref{t:logicalerrors} for each CNOT$_{k-1}$.

If there is one level $k-1$ failure and the ancilla is still accepted, then the state$_k$ must be I, and there can only be one subblock in relative error.  
Indeed, if the single failure is among the transverse Corrections$_{k-1}$ or transverse CNOTs$_{k-1}$, then in the output block$_k$, only one subblock$_{k-1}$ is affected.  
If the failure is in one of the earlier CNOTs$_{k-1}$ or in the preparation of ancillas$_{k-1}$, then all output subblocks$_{k-1}$ will still be well (because the transverse CNOTs succeeded), and also the unverified ancillas will both have well subblocks$_{k-1}$ (because the Corrections$_{k-1}$ succeeded).  It is then possible that there are X or Y relative errors$_{k-1}$, or that the error has spread so there is an X or Y error in the state$_k$ of the unverified ancilla.  In either case, the error will be detected and the ancilla discarded.  

On success, the state$_k$ is either I or Z.  Since we are preparing $\ket{0}_k$, logical Z (transverse physical Z) has no physical effect.  Therefore, if the state$_k$ is Z, we can as a bookkeeping reduction add Z errors to every bit to reduce the state to I.  For this reason, we also do not need to check for Z relative errors.  (Checking for Z relative errors might be a useful optimization -- see Sec.~\ref{s:estimates} -- but it is very intuitive anyway that X errors are worse than Z errors in $\ket{0}_k$ ancillas prepared for X error correction as in Claim B$_k$ \cite{Steane02,Steane03}.)
  
Therefore, on success the output is well$_k$($b_1,\ldots,b_{k-1},a_k$), with 
\begin{equation*}
a_k \equiv \frac{ 2n (A_{k-1} + B_{k-1}) + (2s(n)+n) C_{k-1} }{ (1-A_k) N_k } . \qedhere
\end{equation*}
\end{proof}

\section{5-qubit code distillation} \label{s:moregeneralCSS}

\begin{lemmamagic}
Perfect adaptive stabilizer operations together with the ability to prepare states $\rho_i$, $i = 1, 2, \ldots$, with Bloch sphere coordinates $x_i = y_i = z_i = f_i / \sqrt{3}$, such that $f_i \geq \sqrt{3/7} + \epsilon$, $\epsilon > 0$, gives quantum universality.  Here $\epsilon > 0$, a constant, is known, but the $f_i$ may not be.
\end{lemmamagic}

\begin{proof}
Take five of the $\rho_i$ and decode the five-qubit code stabilized by $X Z Z X I$ and its cyclic permutations.  Postselect on no detected errors.  Bravyi and Kitaev \cite{BravyiKitaev04} show that when the input $f_i$ are all equal to $f$, the output has coordinates $x=y=z = -f_\text{out} / \sqrt{3}$, with $f_\text{out} > f$.  Repeating the postselected decoding procedure efficiently moves the output towards $\ket{T}$, a ``magic" state for which a simple argument gives universality.  A lower bound on $f-\sqrt{3/7}$ is required in order to know how many distillation iterations are required, but $f$ itself need not be known.  

It remains to show that distillation also succeeds when the $f_i$ are not all equal, but only $f_i \geq f \geq \sqrt{3/7} + \epsilon$.  Simple algebra gives that $\partial f_\text{out} / \partial f_i > 0$, so improving any of the input fidelities can only improve the output fidelity.  The probability of decoding acceptance is also monotone, so distillation remains efficient.  

Indeed, as a function of $f_1, \ldots, f_5$, the output fidelity conditioned on decoding acceptance is
$$
f_\text{out} = \frac{(f_1 f_2 f_3 + \cdots) - 2 f_1 f_2 f_3 f_4 f_5}{3 + (f_1 f_2 f_3 f_4 + \cdots)} ,
$$
where ellipses indicate symmetrical terms (in the numerator, nine products of three $f_i$; in the denominator, four products of four $f_i$). 
Differentiate with respect to $f_5$ -- other derivatives are related by symmetry.  Use the quotient rule $d (a/b) = \tfrac{1}{b^2}(b\, da - a\, db)$.  The denominator is clearly at least nine for $f_i \geq 0$.  
As notation, write $f_{i j}$ for $f_i f_j$ and similarly $f_{i j k}$ and $f_{i j k l}$ for three- or four-wise products.
The numerator, which does not involve $f_5$, is 
\begin{gather*}
3 (f_{1 2} + \cdots) - (f_{1 2 3}^2 + \cdots) - (f_{1 2}^2 f_{3 4} + \cdots) - 6 f_{1 2 3 4} - 2 f_{1 2 3 4}^2 \\
= f_{1 2} \left( 3 - f_{1 2 3 4} - f_{3 4} - \tfrac{1}{3} f_{1 2} f_{3 4}^2 - \tfrac{1}{3} f_{1 2}(f_3^2 + f_4^2) \right) + \cdots .
\end{gather*}
Each term is nonnegative when the $f_i \in [0,1]$, implying that $f_\text{out}$ is monotone in each $f_i$ separately. 

The probability of decoding acceptance, 
$
\tfrac{1}{48}(f_1 f_2 f_3 f_4 + \cdots) 
$,
is also monotone.
\end{proof}

\end{document}